\documentclass[aip,rsi,twocolumn,graphicx,caption,superscriptaddress,11pt]{revtex4-1}
\usepackage{mhchem,graphicx,extdash,textcomp,upgreek}
\begin{document}
\title{High-Pressure Laser Floating Zone Furnace}
\author{Julian L. Schmehr}
\email[]{jschmehr@ucsb.edu}
\affiliation{Materials Department, University of California, Santa Barbara, California 93106, USA}
\author{Michael Aling}
\affiliation{Materials Department, University of California, Santa Barbara, California 93106, USA}
\author{Eli Zoghlin}
\affiliation{Materials Department, University of California, Santa Barbara, California 93106, USA}
\author{Stephen D. Wilson}
\email[]{stephendwilson@ucsb.edu}
\affiliation{Materials Department, University of California, Santa Barbara, California 93106, USA}

\begin{abstract} 
 The floating zone technique is a well-established single crystal growth method in materials research, able to produce volumetrically large specimens with extremely high purities. However, traditional furnace designs have relied on heating from high-powered bulb sources in combination with parabolic mirrors, and hence are constrained to transparent growth chambers with large solid angles of optical access. This results in a stark limitation on achievable processing gas pressures, and in turn renders a range of compounds unsuitable for crystal growth by the floating zone technique, either due to excessive volatility or due to metastability. Here, we demonstrate a novel high-pressure laser-based floating zone system (HP-LFZ). The use of lasers for heating allows implementation of a high-strength metal growth chamber, permitting greatly enhanced processing pressures over conventional mirror-based designs, with the current design allowing for pressures up to 1000~bar. We demonstrate a series of example single crystal growths using this design in pressures up to 675~bar, a significant increase over processing pressures attainable in commercially available floating zone systems. The general utility of the HP-LFZ is also illustrated via growths of a range of complex oxides.
\end{abstract}

\maketitle

\section{Introduction}
The availability of pristine single crystals is essential to the discovery of new physical phenomena in condensed matter physics. Sample purity is key in the observation of delicate electronic phases easily quenched or obscured by disorder, while large sample volumes allow for a larger array of experimental techniques such as neutron scattering and other weakly interacting probes to be leveraged in experimental studies.  In the quest to produce ever-cleaner samples of a wide range of materials, an extensive array of crystal growth techniques have been developed~\cite{Schmehr_2017}. Of these, the floating zone technique is particularly impactful~\cite{Pfann_1952,Keck_1953,Theurer_1962}. Here crystal growth occurs out of a molten zone suspended purely by the surface tension between feed material and a seed crystal, rendering the technique crucible-free. The floating zone growth mechanism further facilitates purification via zone refinement of crystals pulled from the melt.

The most common floating zone furnaces use an optical heating source with one or more high-power light bulbs focused into a concentrated volume by parabolic mirrors.\cite{koohpayeh2008optical} For congruently melting materials, an upper feed rod comprised of polycrystalline material is then suspended above a seed rod or crystal in a vertical configuration. A union is formed by melting the tips of both rods and fusing them together, establishing a \textit{floating} molten zone. By slowly translating both feed and seed rods downward, crystal growth is initiated at the seed crystal liquidus line and the volume of the molten zone is controlled by the feed rod's descent into the melt.  Both feed and seed material are moved via translation shafts that seal an inner growth chamber where the gas environment can be controlled.

Employing focused, incoherent light from halogen bulbs or xenon lamps as a heating source requires a large solid angle of optical access for this inner growth chamber.  As a result, the chamber is constructed out of an optically transparent material, usually quartz or sapphire. This introduces a limitation in the mechanical strength of materials that the growth chamber can be built from, which in turn limits the gas pressures that can be applied during growth. The pressure limitation has traditionally restricted which compounds can be grown via floating zone. For instance, outgassing/volatility inherent to some materials becomes prohibitively large as their melting points are approached and growth can only be stabilized by applying large processing gas pressures during growth (\emph{e.g.} providing large oxygen partial pressures).

 Typical high-pressure configurations of commercial floating zone systems utilize thick quartz growth chambers with standard 10 bar pressure limits. The state-of-the-art, commercially available high-pressure floating zone systems produced by SciDre~GmbH, Germany, have pushed this boundary further by employing a two-mirror vertical arrangement in which a single high-power xenon light bulb is located below the growth chamber~\cite{Balbashov_1981,Souptel_2007}. This design allows for the introduction of a thick sapphire processing chamber that increases achievable pressures to 300~bar.

A recently-developed alternative to relying on large areas of optical access for floating zone growth is to instead use laser-based heating. An example of this is the commercially-available laser-based floating zone system produced by Crystal Systems Corp.~\cite{ito_2013}, that employs light pipes to channel light from five 200~W diode lasers into the growth zone. This arrangement allows for extremely sharp axial heating gradients (exceeding 150 \textdegree{}C/mm), high circumferential uniformity and elevated growth temperatures; however the design still relies on a quartz growth chamber with limited processing pressures.

Here, we describe the design and construction of a new high-pressure laser floating zone (HP-LFZ) furnace. Heating power is supplied by a set of continuous diode lasers operating in a wavelength range optimized for an array of oxides and with exceptional power stability ($<0.1$\% drift in output power over a 7 day time span). Conventional optics is used to focus the laser beams onto a sample and allow for the introduction of a metal growth chamber with discrete points of optical access provided by high-pressure windows. This substantially increases the maximum processing gas pressure possible for floating zone crystal growth to 1000~bar, a more than three-fold increase relative to the current state-of-the-art. The high processing gas pressures available in the HP-LFZ furnace, coupled with the proven effectiveness of laser-based heating \cite{ito_2013}, opens a wide range of new compounds to growth by the floating zone technique.

\section{Instrument Description}

\begin{figure*}
\centering
\includegraphics[width=\textwidth]{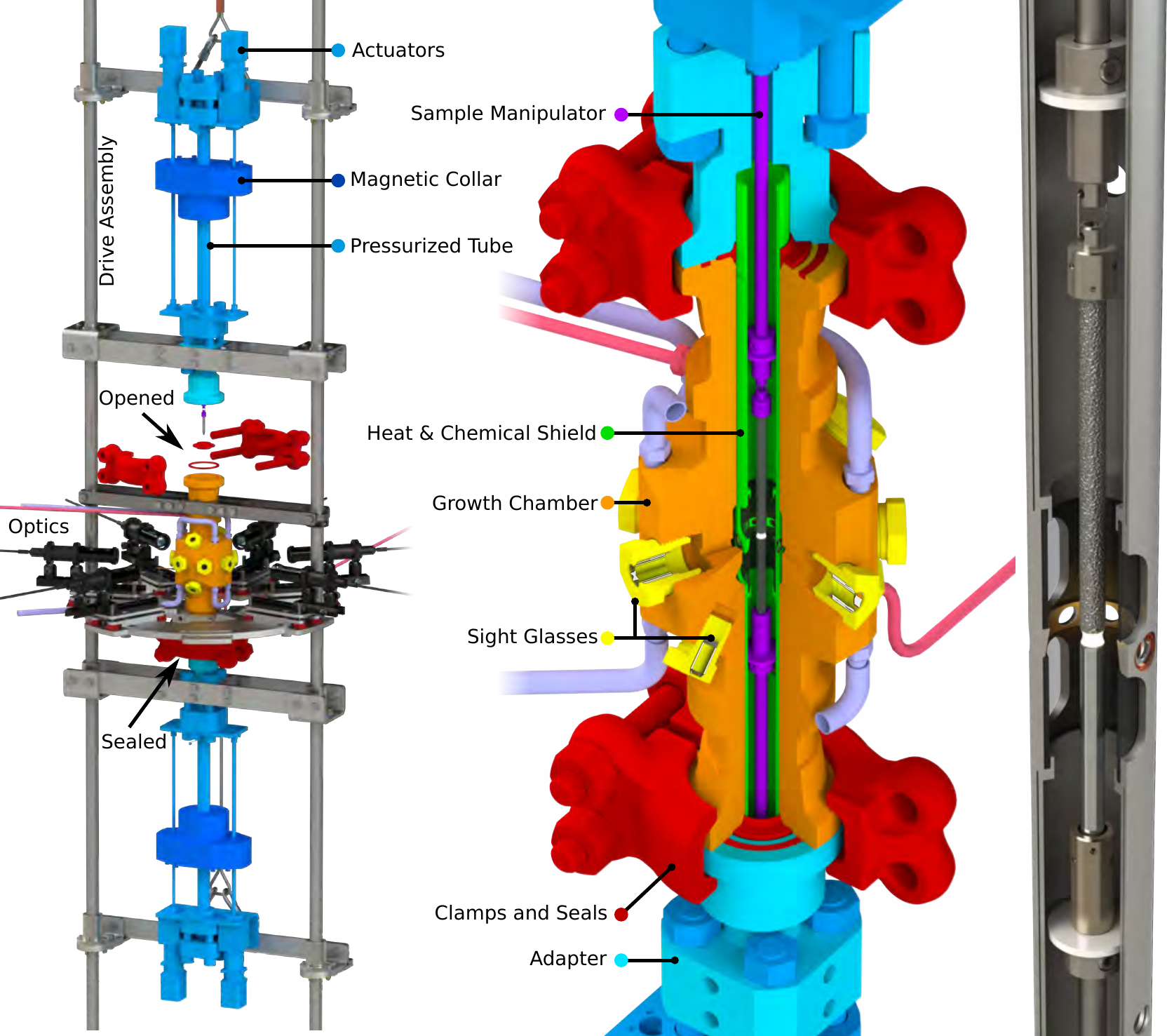}
\caption{\label{fig:colorview}Exploded view of overall assembly (left) and cut view of core components (center). The exploded view is a partial illustration of the process for sample mounting and alignment, while the cut view illustrates conditions during growth. A detailed rendering (right) shows the suspension of the feed rod (top) and seed rod (bottom), connected by the molten zone and surrounded by the heat and chemical shield. Alumina baffles (white) shield the growth chamber along its axial extent, and mitigate convection within the growth chamber.}
\end{figure*}

\subsection{Growth Chamber}
A color-coded overview of the instrument is shown in Fig. \ref{fig:colorview} (color online).  A metal growth chamber contains the pressurized growth environment within a vertical central bore ($330$ mm long and $25.4$ mm in diameter.) Seven equally spaced, 5 mm diameter in-plane bores allow for laser access to the sample growth position. Four 5 mm diameter out-of-plane bores allow for optical observation of the growth process, and narrower (2 mm diameter) bores are used for gas inlet and outlet. The chamber was fabricated from the structural alloy UNS~N07718 (Inconel~718), which is a nickel-base superalloy with exceptional fracture toughness, stable high-temperature mechanical properties, and intrinsic resistance to chemical attack and oxidation. All metal components are designed for sustained operation at 1000~bar and simulations show that they are tolerant to surface cracks up to several millimeters at that pressure. This is in contrast to existing designs for high-pressure floating zone growth chambers, where the brittleness of a quartz or sapphire chamber enables rapid crack propagation. 

Eleven sight glasses (seven laser ports and four observation ports) fabricated by Encole~LLC provide optical access to the sample environment. Each of these is a bonded construction of anti-reflection-coated fused silica inside a UNS~S17400 (17\Hyphdash* 4PH) stainless steel housing, with an 11 mm clear aperture at the outer face. The sight glasses' steel housings feature tapered NPT pipe threads which are lubricated by an oxygen-compatible PTFE tape and mate to the central growth chamber in a replaceable manner. Channels machined within the growth chamber allow for liquid cooling from a water-based chiller, which maintains chamber external temperatures below 50 \textdegree{}C. Three thermocouples embedded at varied positions and depths in the chamber walls monitor the chamber temperature. 

\begin{figure*}
\centering
\includegraphics[width=\textwidth]{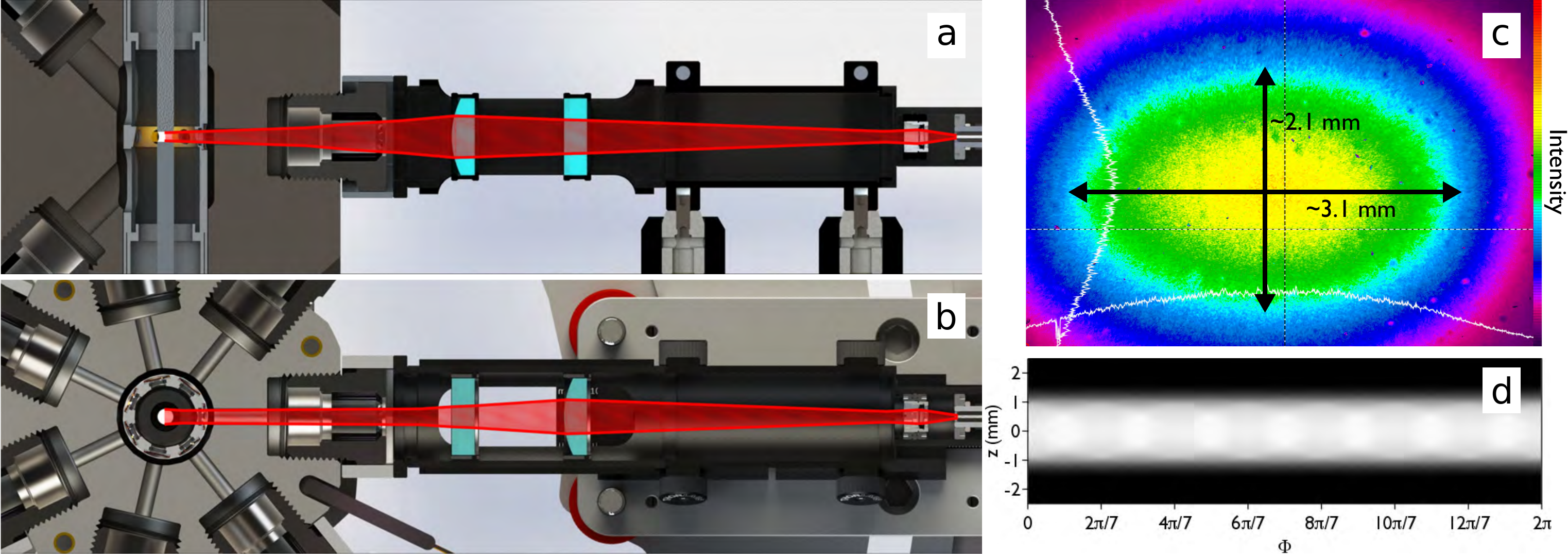}
\caption{\label{fig:beams}Depiction of the optical lens array focusing light into the growth chamber.  Plano-convex cylindrical lenses enable independent focusing in the vertical (a) and horizontal (b) directions. This results in an oblong spot at the sample position, imaged in (c). The arrows indicate approximate Gaussian FWHM in the horizontal and vertical directions, white lines along the axes show the intensity distribution across the spot profile in the horizontal and vertical directions. For an approximate visualization, the simulated superpositon of seven laser spots on a 4~mm diameter cylindrical sample is shown in (d).}
\end{figure*}

A separate, removable heat and chemical shield (see right panel of Fig. \ref{fig:colorview}) lines the central bore of the growth chamber and isolates it from direct thermal radiation, ejected molten material, and volatilized powder during crystal growth. The central shield housing is made from an aluminum alloy. Two replaceable quartz cylinders allow continued observation through the viewing ports, and seven anti-reflection coated sapphire rounds nested inside the shield's wall allow for laser light transmission. An angled inner surface at the shield's midplane reflects stray laser light out of the optical plane of the lasers and reduces the intensity reflected back through the chamber's outer windows. The shield's aluminum body also acts to distribute heat away from the chamber core near the molten zone.

\subsection{Laser Heating and Optics}

Optical heating power is provided by seven 100~W Coherent Inc$.$ HighLight~FAP diode laser modules, which provide a combined output power of 700~W at 810~nm \Emdash a wavelength at which many oxides exhibit high absorptivity. The lasers are fiber-coupled, and their light is transported through Coherent 830\Hyphdash* $\upmu$m fibers to optical lens tubes arrayed radially about the chamber's central axis. The lens tubes contain a collimating lens and focusing optics.  They use plano-convex cylindrical lenses to apply independent horizontal and vertical focusing of the collimated fiber output. A typical laser spot size, as determined by its Gaussian full width at half-maximum (FWHM), is as narrow as 2.1~mm vertically but is typically focused to be slightly broader, 3.1~mm, in the horizontal direction. An image of the beam spot, imaged using an Ophir LBS-300 beam splitter and Newport LBP2 beam profiler, is shown in Figure \ref{fig:beams}~(c). White lines along the axes show the intensity distribution across the spot profile in the horizontal and vertical directions. The superposition of these oblong spots creates a smooth heating profile around the molten zone, with sharp thermal gradients in the growth direction (see Fig. \ref{fig:beams} (d) for a simulation of the light intensity on a 4 mm diameter cylindrical sample). 

Still images and videos of crystal growth are captured by a Thorlabs Inc. Quantalux monochrome camera and a series of imaging lenses through one of the out-of-plane chamber sight glasses. Another out-of-plane sight glass is used to allow a Fluke Process Instruments Endurance two-wavelength pyrometer to measure molten zone temperatures and capture secondary images. High-optical-density notch filters centered at 800~nm absorb laser light reflected through the instrumentation sight ports before it can reach the camera or pyrometer. The two remaining, auxiliary optical ports can be used to further monitor temperature and growth conditions or to allow for the use of other optical probes of the growth zone.
\begin{figure*}[ht]
\centering
\includegraphics[width=\textwidth]{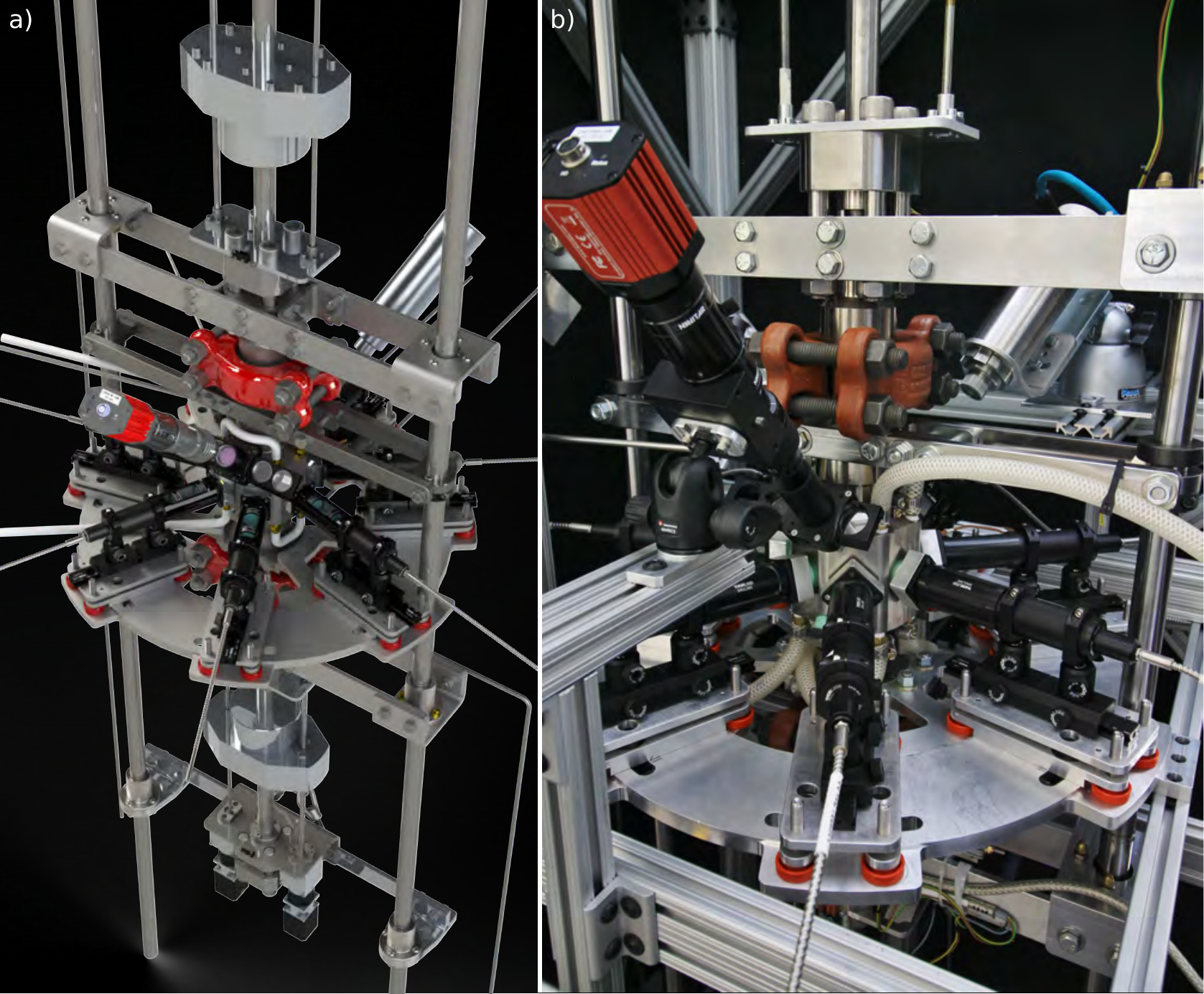}
\caption{\label{fig:overview}(a) Computer-generated rendering of full apparatus with all components other than the external frame shown. In addition to the elements shown in Fig. \ref{fig:colorview}, the camera (foreground, red and gray finish), pyrometer (background, metallic), and their filters are visible. (b) For comparison, a photograph taken of the central region of the actual apparatus, configured as it would be during a growth.}
\end{figure*}

\subsection{Translation system}

Smooth translation of both the feed and seed material is vital for establishing a stable crystal growth. In traditional floating zone systems, motion is achieved by driving a shaft through an o-ring or magnetic fluid pressure seals. At elevated pressures such operation becomes extremely challenging because of the large frictional forces involved. Maintaining a seal against 1000~bar internal pressure while allowing for smooth motion of the translation shafts through dynamic seals with negligible slipping was therefore not pursued. Instead, the HP-LFZ design makes use of magnetically coupled drives. Here the drive shaft is completely enclosed in the high-pressure environment and coupled to an external drive system via a magnetic collar. This eliminates the need for a dynamic seal and high torque drive motors.

The drive system developed by SciDre~GmbH was adopted to both translate and rotate the shafts driving to the feed and seed rods during crystal growth. The coupling between an external magnetic collar and an internal shaft fully enclosed within the pressurized environment reduces the load experienced by the actuators, thereby removing otherwise severe backlash effects and unwanted frictional slip. These drives are sealed to the growth chamber via a pair of adapting collars made from Inconel 718. These adapters are sealed to the drives by a metal-on-metal cone seal, and on the other end, they routinely seal with the main growth chamber via a reusable conical metal seal ring and clamp connector. Simple elastomer seals can also be utilized for low-pressure growths. The furnace can be opened in two locations by sliding each drive system along with its mating adapter vertically away from the chamber along a pair of guide rails.  Each carries along the separated feed material or seed crystal and allows access to the central chamber.

\section{Example Growths}
\label{sec:example}

\subsection{Crystal Growth Demonstration}
\begin{figure}
\centering
\includegraphics[width=\columnwidth]{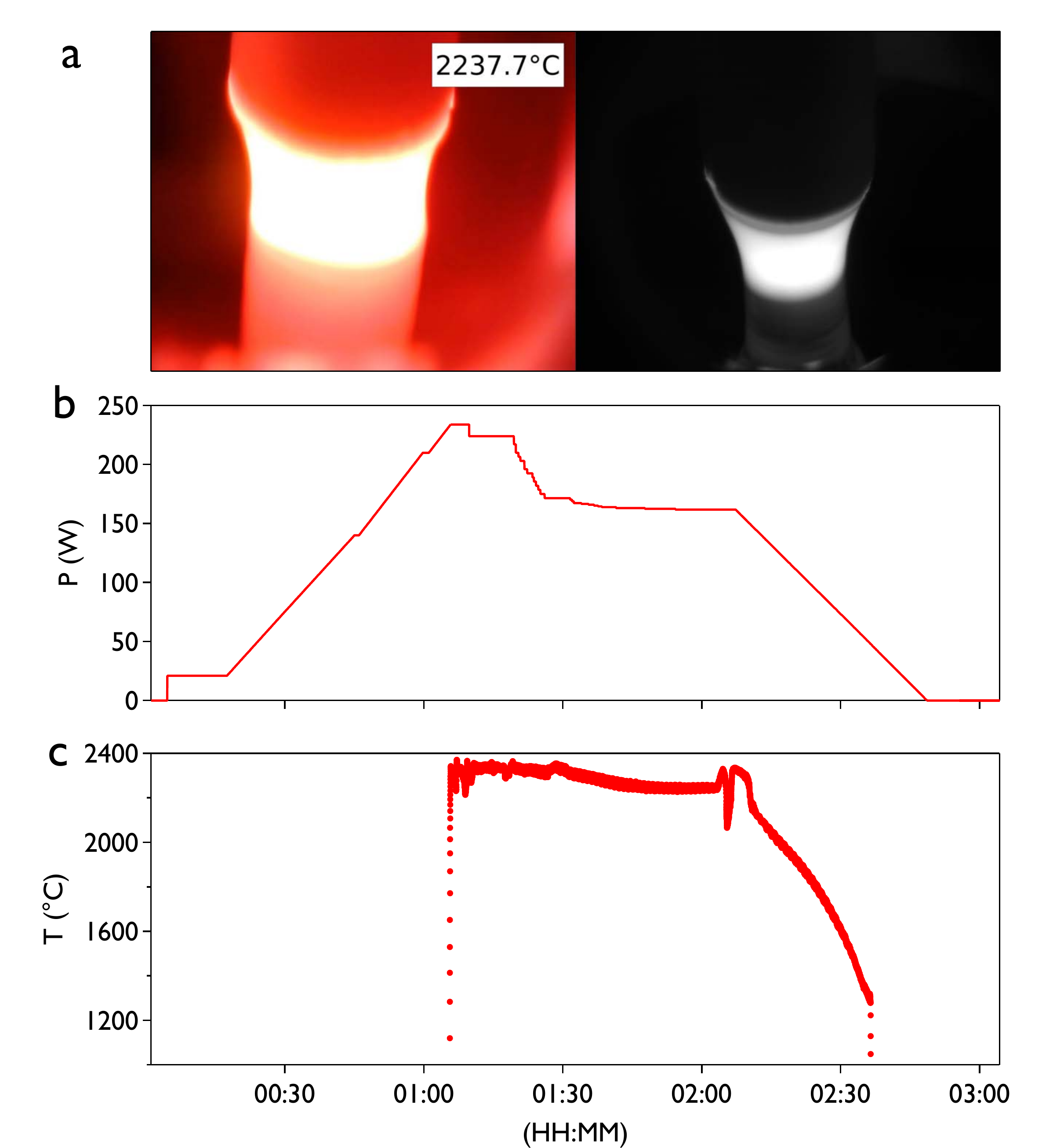}
\caption{\label{fig:Al2O3_growth}(a) Image of the molten zone, taken with the pyrometer camera (left) as well as the Quantalux monochrome camera (rigth) for the growth of \ce{Al_2O_3} in air. (b) Laser power and  (c) measured melt zone temperature as a function of time for the duration of the growth.}
\end{figure}

For basic commissioning, crystals of corundum and ruby (melting point near $\sim$2000\textdegree{}C) were grown at pressures up to 330~bar. \ce{Al_2O_3} powder was ground and pressed into 4~mm diameter feed and seed rods. Ruby powder was prepared by adding 0.5-1\% (molar) of \ce{Cr_2O_3} to the starting powder. The rods were sintered at 1650\textdegree{}C for 36 hours on a bed of \ce{Al_2O_3} powder in order to minimize stress on the sample. Growths were performed at 20~mm/hr with feed and seed rods counter-rotating at 11~rpm and $10$~rpm respectively. Images of the crystal growth of corundum in air are shown in Fig. \ref{fig:Al2O3_growth}~(a).

\begin{figure}
\centering
\includegraphics[width=\columnwidth]{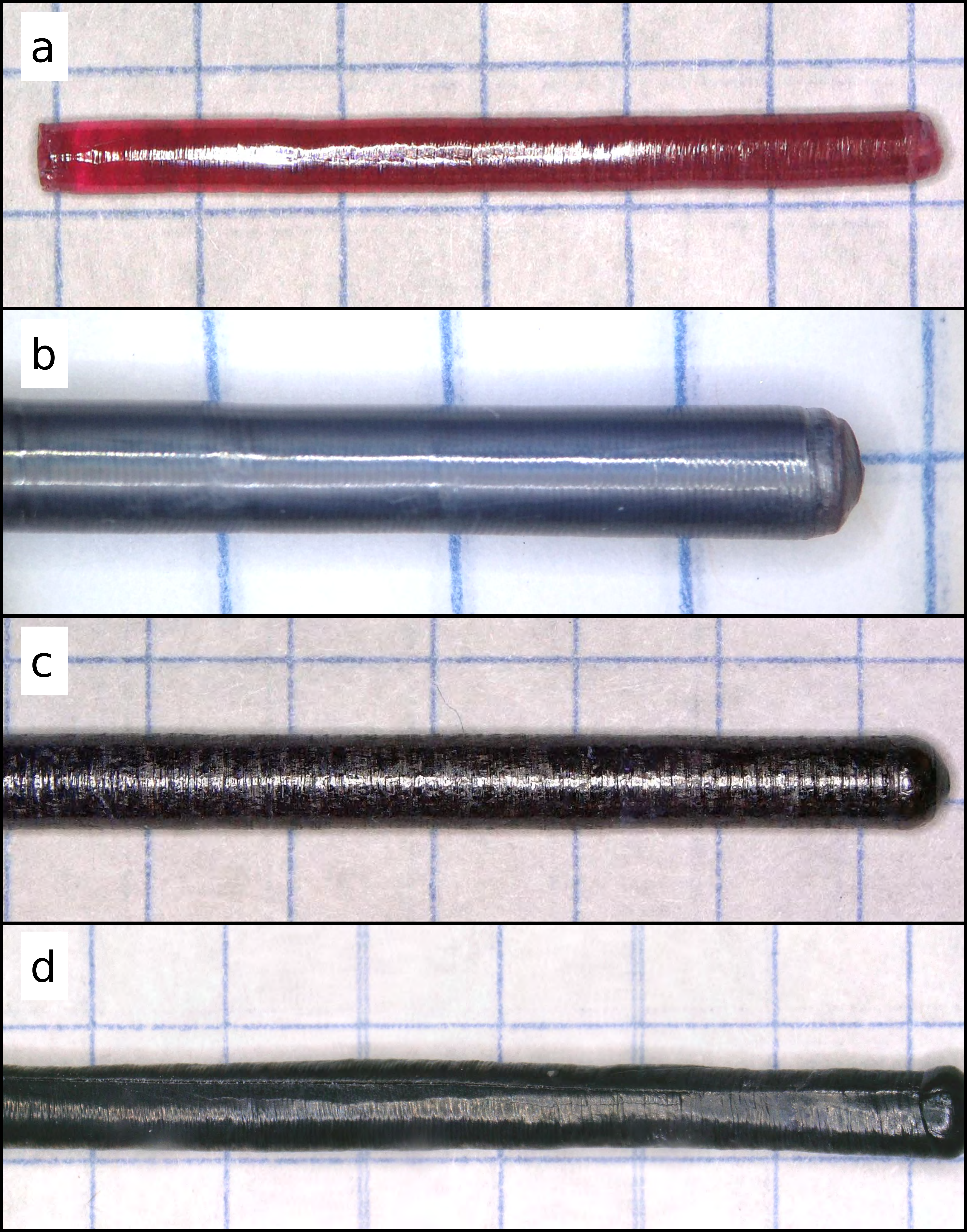}
\caption{\label{fig:sample_images}Example single crystals grown in the HP-LFZ. (a) Ruby grown using 70~bar Ar. (b) \ce{GdTiO_3} grown using 35~bar Ar. (c) \ce{Nd_2Zr_2O_7} grown using 70~bar Ar, (d) \ce{Li_2CuO_2} grown using 105~bar 80:20 \ce{O_2}:\ce{Ar} mixture. The blue grid lines are 5 mm apart.}
\end{figure}

A stable melt zone was attained using between $160$\Hyphdash*$165$~W.  We note here that \ce{Al_2O_3} powder in high purity (5N) form can only be melted via an initial high intensity optical pulse which heats the material sufficiently to activate absorption.  Once molten, the zone can be stabilized at reduced laser power and high purity crystals of this large band gap material grown. The addition of \ce{Cr_2O_3} increases the absorptivity of the sample sufficiently that the initial heating pulse is no longer required. On the other hand, in this case ruby fluorescence causes a significant deviation of the pyrometer's measured temperature at low laser powers. We note here that for both corundum and ruby growths, significant laser transmission through the sample results in a deviation of the measured melting temperature by $\approx200$~\textdegree{}C from the expected values. One example of a ruby crystal grown under a 70\Hyphdash* bar Ar atmosphere is shown in Fig. \ref{fig:sample_images} (a), where a facet can be seen running along the length of the crystal. 

\subsection{High-Pressure Demonstration}

\begin{figure}
\centering
\includegraphics[width=0.9\columnwidth]{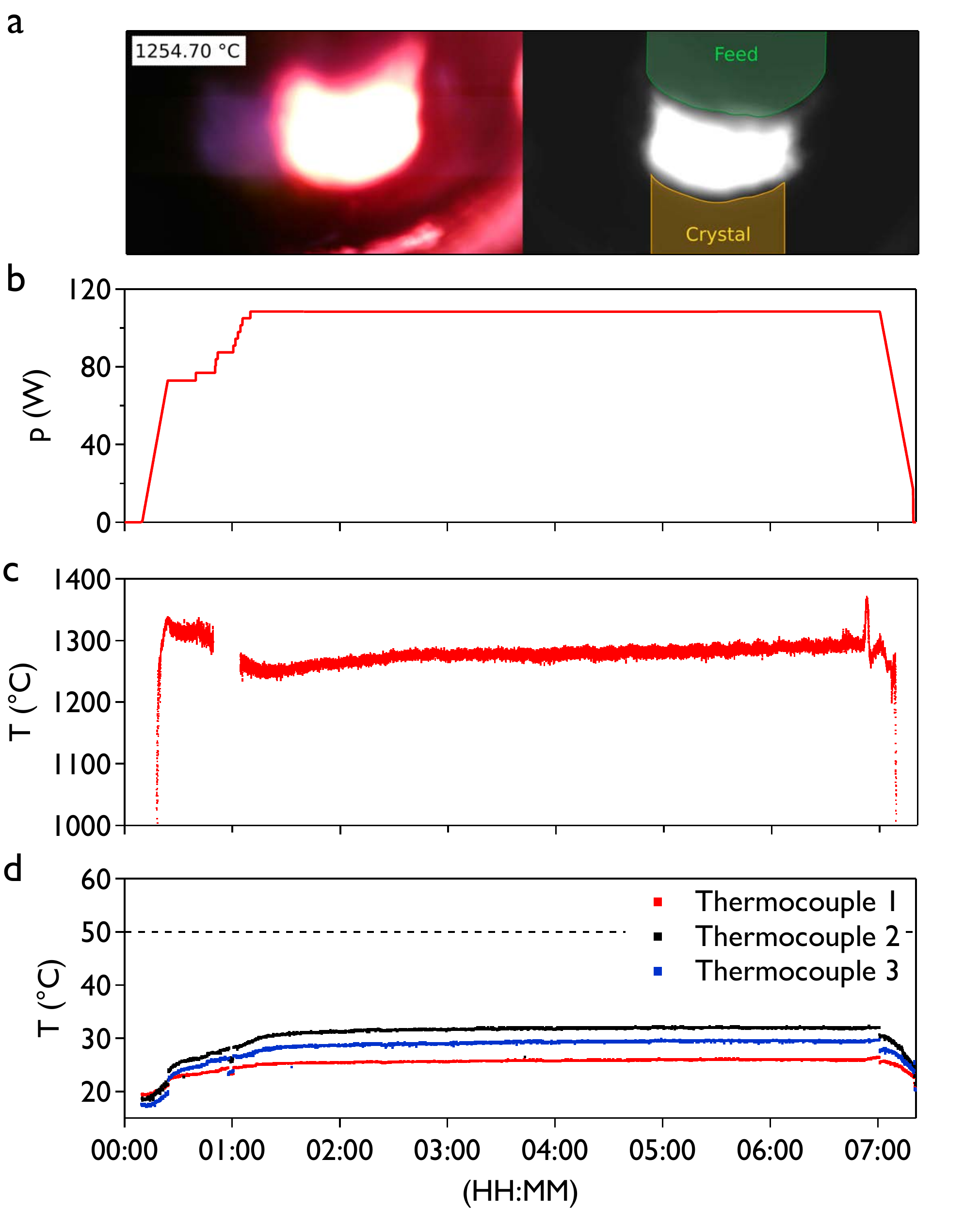}
\caption{\label{fig:Cu2O_growth}(a) Image of the molten zone, taken with the pyrometer camera (left) as well as the Quantalux monochrome camera (right) for the high pressure growth of \ce{Cu_2O}. The elevated temperature inside the chamber leads to convection currents that disturb the image, particularly at high pressures. The sample outlines of feed and seed are highlighted in the camera image. (b) Laser power, (c) measured melt zone temperature, and (d) measured growth chamber temperatures as a function of time during the growth. The gap in measured sample temperature originates from a realignment of the pyrometer during growth.}
\end{figure}

The core capability of the HP-LFZ to stabilize crystal growth at high pressure was demonstrated by the growth of a \ce{Cu_2O} crystal under 675~bar Ar pressure. Starting rods of OFHC copper (3.125~mm diameter, 4N purity) were oxidized in air at 1150~\textdegree{}C for 72~hrs, followed by a quench to reduce CuO formation, using a vertical tube furnace~\cite{Ito_1998}. A short rod ($25$~mm length) was employed as a seed with a longer $75$ mm rod used as the feed. To initiate growth, the total laser heating power was ramped to 110~W, at which point the tip of the \ce{Cu_2O} seed sample became molten (melting point $\approx$1250~\textdegree{}C). The feed rod was then lowered into the melt and the growth was initiated when the established molten zone appeared stable (see Fig.~\ref{fig:Cu2O_growth}).

Crystal growth was performed at 5~mm/hr for a total of 6 hours, with feed and seed shafts counter-rotating at 7~rpm and $6$~rpm respectively. Convection is significant at these elevated pressures, with turbulent flow impacting optical inspection of the melt (see Fig.~\ref{fig:Cu2O_growth}~(a) for camera images of the growth at 675~bar); however this has seemingly little impact on the molten zone itself. Measurements of the heating power, sample temperature and chamber temperatures are shown in Fig.~\ref{fig:Cu2O_growth}~(b-d). Despite severe convection inside the chamber, the molten zone remained stable throughout the growth. Powder x-ray diffraction (XRD) measurements of crushed crystals confirmed the phase purity apart from a small amount of \ce{CuO} stabilized at the surface.

\subsection{Demonstration of Zone Stability }

In order to demonstrate the stability of the molten zone using laser-based heating in a high pressure environment, an oxide with a low viscosity melt, GdTiO$_3$, was grown.  \ce{GdTiO_3}, a collinear ferrimagnet,\cite{greedan1985rare} is on the border of the ferromagnetic-antiferromagnetic transition in the rare earth titanate series (\ce{RTiO3}). High purity, stoichiometric crystals are challenging to grow in conventional mirror furnaces \cite{roth_2008,amow_2000}. This is due to the high density of the melt, which can overwhelm the surface tension and destabilize the molten zone \cite{roth_2008}. In a conventional furnace, the molten zone tends to elongate over time, often leading to separation of the feed and grown crystal. 

To grow this material, powder was initially synthesized based on the procedures detailed in earlier studies.\cite{roth_2008,amow_2000} Starting powders of \ce{Gd_2O_3} (Alfa Aesar, 99.99\%), \ce{TiO_2} (Alfa Aesar, 99.99\%) and Ti (Alfa Aesar, 99.99\%, dehydrided) corresponding to the stoichiometry \ce{GdTiO_{2.93}} were ground together. \ce{TiO_2} and Ti were used due to the stability of the Ti valence in these precursors, while the starting stoichiometry was chosen based on previous thermogravimetric analysis (TGA) of grown crystals \cite{roth_2008}.  This procedure seemingly works best for growth in an Ar environment presumably due to slight oxidation during sintering and growth.  Growth with nominally stoichiometric GdTiO$_3$ starting material instead requires a mixed H$_2$/Ar reducing environment \cite{PhysRevB.56.10145, amow_2000}. 

 Powders were mixed and then pressed at 3100 bar using a cold isostatic press. The pressed powder was then placed on a Mo plate and fired at 1400\textdegree{}C for 18 hrs under 10$^{-8}$ bar\cite{amow_2000}. 4 mm diameter feed and seed rods (100 mm and 50 mm in length, respectively) were then pressed from this powder, hung vertically using Mo wire, and fired at 1200\textdegree{}C for 10 hrs under the same vacuum conditions. The ramp rate was kept at 5\textdegree{}C/min to avoid thermal cracking. 

\begin{figure}
	
\centering
\includegraphics[width=\columnwidth]{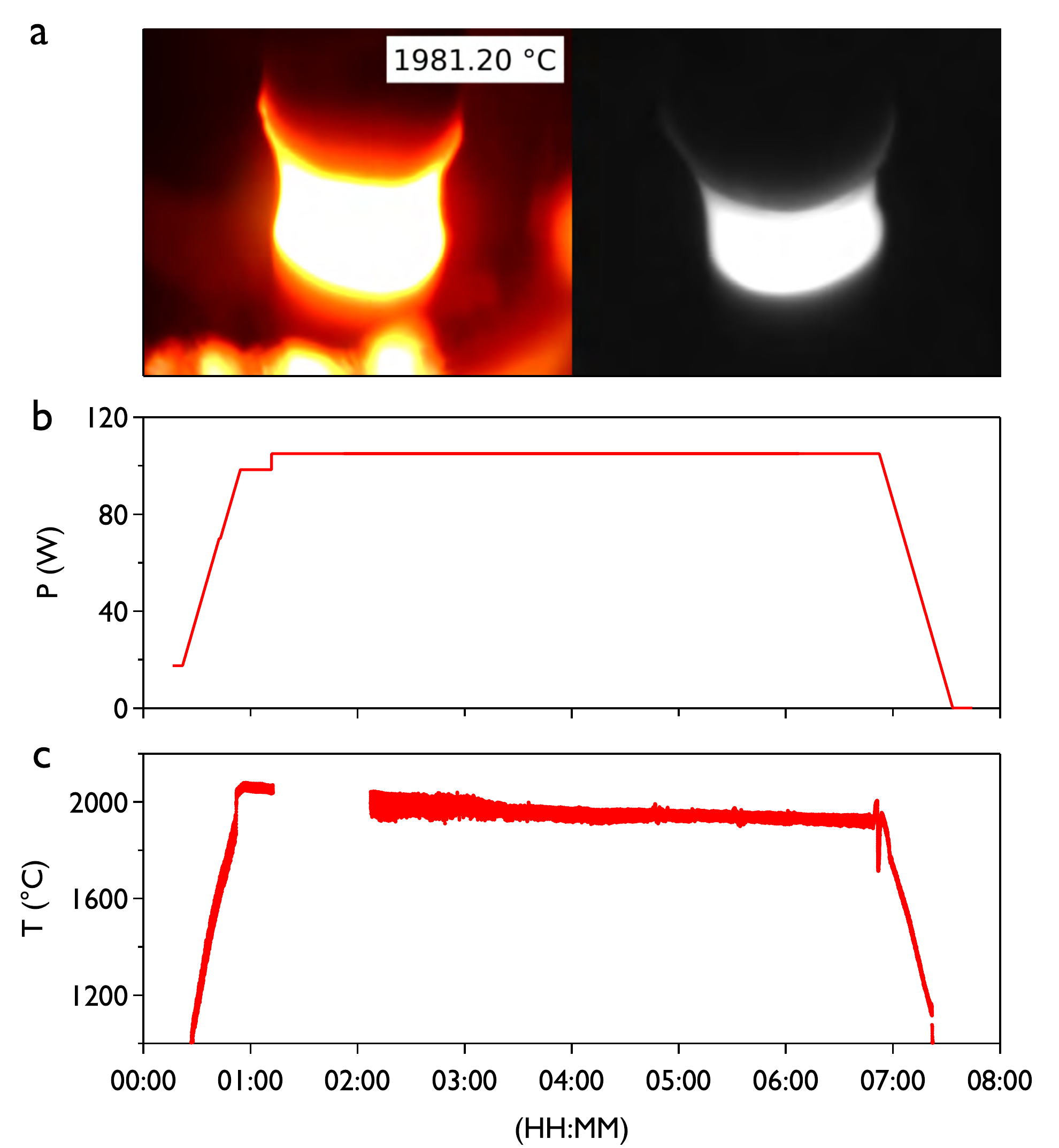}
\caption{\label{fig:GdTiO3_growth}(a) Image of the molten zone, taken with the pyrometer camera (left) as well as the Quantalux monochrome camera (right) for the growth of \ce{GdTiO_3} under 35~bar Ar. (b) Laser power and  (c) measured zone temperature as a function of time. During joining of the feed and seed rods no consistent temperature readings could be taken, leading to the gap in the data in (c).}
\end{figure}

The sharp thermal gradient and tight heating zone of the HP-LFZ, as well as the ability to routinely employ modest atmospheric overpressure, proved easily capable of creating a stable melt of \ce{GdTiO_3}. The growth was performed under static 35~bar ultra-high purity Ar following several cycles of purging and re-pressurizing the chamber to remove residual oxygen. A stable molten zone was established using a total heating power of 105~W, and a 50 mm long crystal was pulled at a rate of 10~mm/hr. Images taken during the growth, as well as the growth parameters and zone temperature are illustrated in Fig.~\ref{fig:GdTiO3_growth}.

\begin{figure}
\centering
\includegraphics[width=\columnwidth]{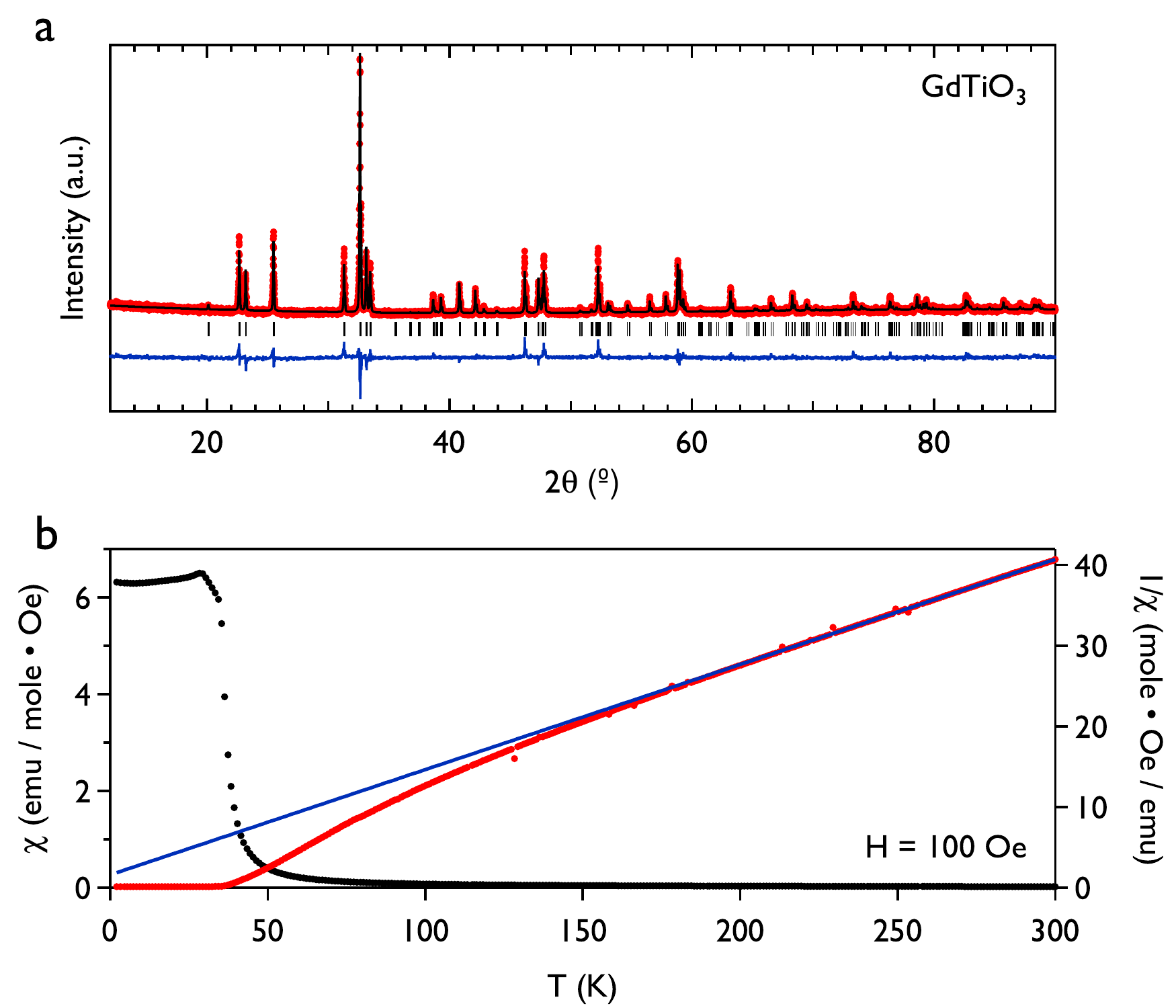}
\caption{\label{fig:GdTiO3} Characterization of \ce{GdTiO_3} crystal grown in the HP-LFZ furnace. (a) powder X-ray diffraction (PANalytical, Empyrean, Cu $K_{\alpha}$ radiation) on a crushed piece of crystal. All observed peaks can be indexed to the reported \ce{GdTiO_3} structure (black tick marks). (b) SQUID magnetometry (Quantum Design, MPMS 5XL) on a 20 mg piece of the crystal. The Curie-Weiss fit to $1/\chi$ (blue line) yields $\theta_{CW}=-12$~K and the ferrimagnetic T$_c$ was determined to be 35 K.}
\end{figure}

The melt remained stable throughout the growth, and a large facet parallel to the (001)-axis is visible along the length of the crystal shown in Fig. \ref{fig:sample_images} (b). The phase purity of the sample was confirmed by powder x-ray diffraction measurements (Fig. \ref{fig:GdTiO3} (a)), which showed no \ce{Gd_2Ti_2O_7} or other impurities. Lattice parameters were refined using the FullProf package~\cite{Rodriguez-Carvajal_1993} ($R_p$ = 3.59\%, $\chi ^2$ = 2.53) and are consistent with previous reports \cite{amow_2000,roth_2008,zhou_2005}. The crystal quality was further checked via a texture scan on a Panalytical MRD PRO Materials Research Diffractometer, which showed that the facet oriented along the (001)-axis reflected only a single, resolution-limited grain.  We note that there is a consistent increase in the $b$ and $c$ lattice parameters for samples with higher \ce{T_C} \cite{roth_2008,komareK_2007} as compared to those with lower \ce{T_C} \cite{amow_2000,zhou_2005}. Magnetization measurements plotted in Fig.\ref{fig:GdTiO3} (b) show behavior consistent with previous reports\cite{amow_2000,roth_2008,zhou_2005}. Notably, the high \ce{T_C} of 35~K is an indication of near ideal oxygen stoichiometry.

\subsection{High-temperature crystal growth}

One of the advantages of a laser-based heating source is the ability to easily melt refractory materials with melting points traditionally accessible only via compact xenon light sources in conventional mirror-based furnace designs.  As a demonstration of this, a series of successful single-crystal growths of the refractory zirconate \ce{Nd_2Zr_2O_7} demonstrate the high-temperature capabilities of the HP-LFZ. This compound is of interest to a variety of communities as both a thermal barrier coating material as well as an unusual quantum magnet that shows signs of magnetic fragmentation ~\cite{Lhotel_2015,Petit_2016,Lhotel_2018}. A variation exists in the lattice parameters reported for single crystals, grown in ambient air, compared with pristine powder samples, with the lattice parameter of the single crystals being reduced by 0.5\%~\cite{Subramanian_1983,Hatnean_2015}. This potentially indicates a concentration of vacancies and/or off-stoichiometry in crystals grown under ambient pressure and atmosphere, and high pressure growth of this compound in an Ar environment was investigated.

Seed and feed rods were prepared using a conventional solid-state synthesis method\cite{Hatnean_2015}. Stoichiometric mixtures of \ce{Nd_2O_3} (99.99\%, Alfa Aesar) and \ce{ZrO_2} (99.98\%, Alfa Aesar) were mixed using a mortar and pestle and pressed into pellets under 3500~bar hydrostatic pressure. The pellets were fired for two days each at 1300~\textdegree{}C, 1375~\textdegree{}C and 1450~\textdegree{}C, with intermediate grindings. The resulting powder was found to be phase pure with clear superlattice peaks resulting from the pyrochlore structure. The powders were then pressed into seed ($\sim$25~mm length) and feed (50 \Endash~ 100~mm length) rods, which were sintered on a bed of \ce{Nd_2Zr_2O_7} powder at 1450\textdegree{}C for 24~hrs.

\begin{figure}
\centering
\includegraphics[width=\columnwidth]{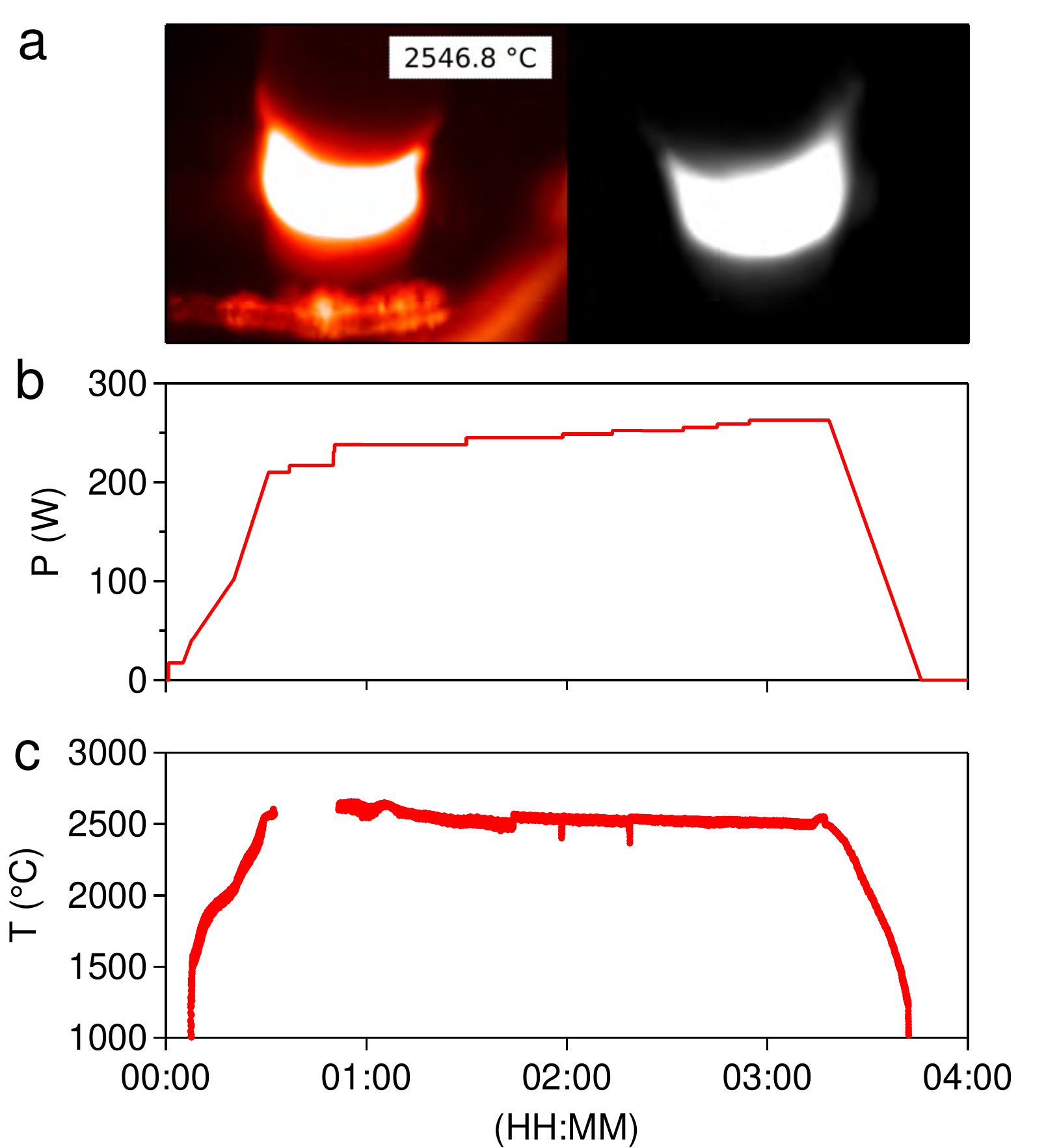}
\caption{\label{Nd227_growth}(a) Image of the molten zone, taken with the pyrometer camera (left) as well as the Quantalux monochrome camera (right) for the growth of \ce{Nd_2Zr_2O_7} in 70~bar Ar. (b) Laser power and  (c) measured melt zone temperature as a function of time. During joining of the feed and seed rods, no consistent temperature readings could be taken, leading to the gap in the data in (c).}
\end{figure}

We performed three successful growths of this material: in ambient air, in 35~bar Ar and in 70~bar Ar growth conditions. Despite the high melting point of \ce{Nd_2Zr_2O_7} (2200~\textdegree{}C), as well as convective heat losses due to the pressurized growth conditions, these growths could be performed under a relatively low level of power, 263~W total (see Fig. \ref{Nd227_growth} for images taken during the growth as well as growth power and measured temperature).  For all three samples, single crystal specimens were successfully grown, indicated by large, parallel facets forming on opposing sides of the grown crystal (see Fig. \ref{fig:sample_images}(c) for a sample grown in 70~bar Ar). Significant deposition of reagents on the chemical shield's disposable windows occurred and this caused a gradual reduction in the transmitted heating power, which was compensated by gradually increasing the laser power (see Fig.~\ref{Nd227_growth}). At slow growth speeds ($<20$~mm/hr), severe cracking of the grown crystal was observed due to sharp heating gradients; however the growth of these cracks can be mitigated by increasing translation speed. This yielded single crystals with only localized fracturing.  Broadly focused lasers, aimed at the growth region immediately below the melt zone through the lower out-of-plane windows, are a potential future step for mitigating thermal stress during growth.  This can emulate ``after-heater" configurations in conventional lamp-based optical furnaces.  

Crystals grew with facets aligned along the (111)-axis and texture scans on an x-ray diffractometer showed only a single grain. The lattice parameters of all three samples and a powder sample from literature are shown in Table~\ref{tab_Nd227_latparam}. The air-grown sample has a lattice parameter comparable to samples grown in similar conditions\cite{Hatnean_2015}, reduced compared to the powder specimens\cite{Subramanian_1983}. However, increasing Ar pressure resulted in a gradual increase in the lattice parameter $a$, and, for 70~bar Ar pressure, the lattice parameter recovers the powder value. The high growth pressure likely limits volatilization during growth and results in reduced vacancy levels and site-mixing in the crystals. A crushed crystal of Nd$_2$Zr$_2$O$_7$ grown under 1000 psi Ar showed no impurity phases with refinement parameters of $R_p$ = 4.87\% and $\chi ^2$ = 1.89, and the crushed Nd$_2$Zr$_2$O$_7$ crystal grown in ambient air refined with parameters $R_p$ = 5.58\% and $\chi ^2$ = 1.48.

 \begin{table}
 \caption{\label{tab_Nd227_latparam} Comparison of \ce{Nd_2Zr_2O_7} lattice parameters under varying growth conditions. P: powder. SC: Single crystal.  
 }
\begin{ruledtabular}
\begin{tabular}{ c c c c}
  Source & Sample &Atmosphere & $a$ (\AA)\\\hline
~\cite{Subramanian_1983} & P & air & 10.678\\
~\cite{Hatnean_2015}  & SC & air &10.62652(9) \\
this work  & SC & air & 10.6240(1)\\
this work   & SC & 35~bar Ar & 10.66387(4)\\
 this work  & SC &70~bar Ar & 10.67171(5)\\  
\end{tabular}
\end{ruledtabular}
\end{table}

\subsection{Growth of Volatile Oxides}

We also performed HP-LFZ growths testing the growth of a known volatile oxide, \ce{Li_2CuO_2}; a material studied extensively due to its quasi-1D structure comprised of $S=1/2$ spin chains. Oxygen off-stoichiometry in this system is suggested to induce a magnetic moment on the oxygen site, in turn leading to a net ferromagnetic magnetization at low temperature.\cite{shu_2017} This becomes a sensitive metric for sample stoichiometry. A substantial oxygen partial pressure (\ce{P_{O2}}) is known to be required for successful growth of \ce{Li_2CuO_2}, and it has been demonstrated that increased \ce{P_{O2}} reduces oxygen off-stoichiometry \cite{shu_2017}. To explore this, we prepared feed rods by solid state synthesis~\cite{Wizent_2011} from a mixture of \ce{LiOH} and \ce{CuO}, with 10\% (mass) excess of LiOH over a stoichiometric mixture. The powders were ground together and prepared for pressing in an Ar-filled glove box, then fired at 750\textdegree{}C for 24~hrs in air. The dry powders were then pressed into rods, placed on a bed of reacted powder and sintered at 750\textdegree{}C for 36~hrs. 

\begin{figure}
\centering
\includegraphics[width=\columnwidth]{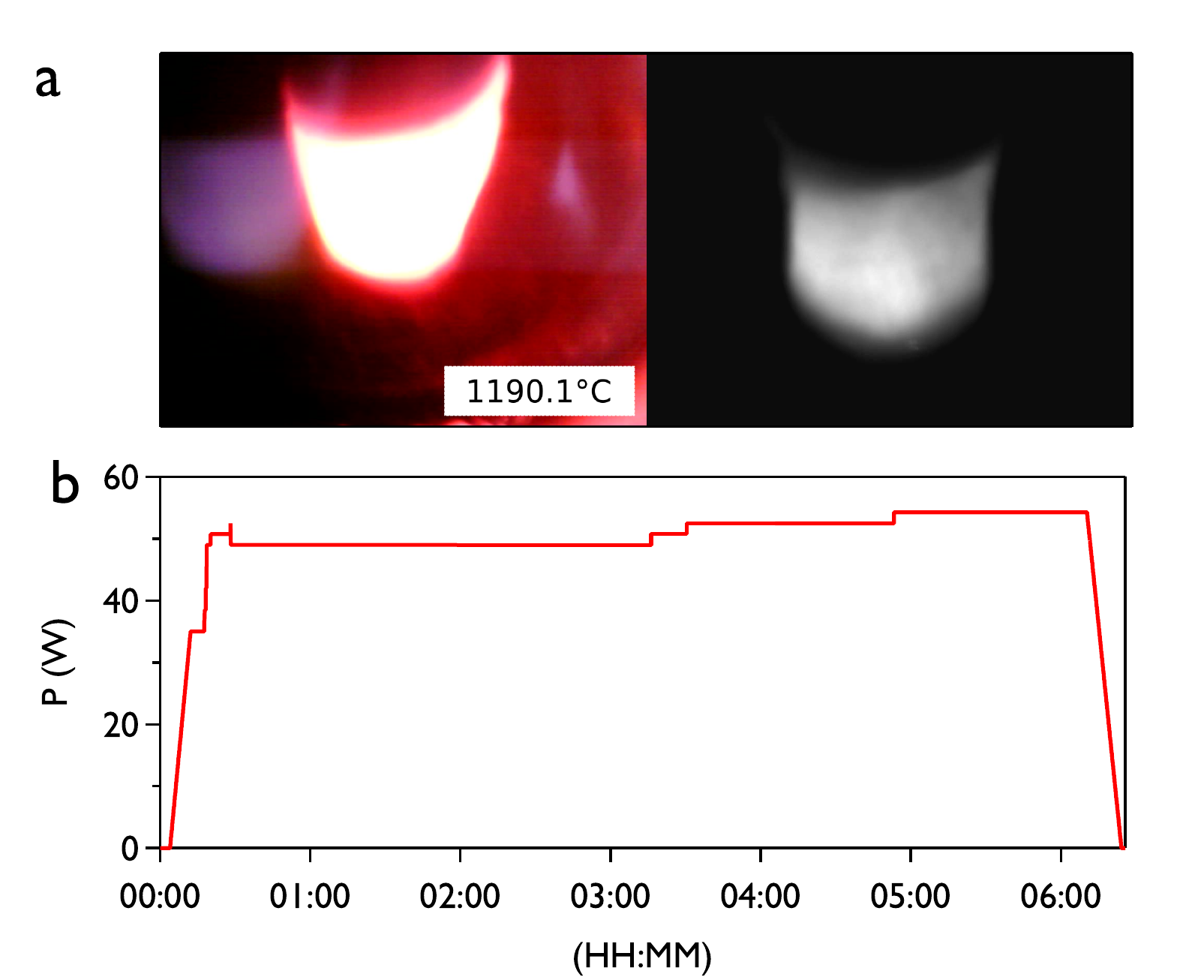}
\caption{\label{fig:Li2CuO2_growth}(a) Image of the molten zone, taken with the pyrometer camera (left) as well as the Quantalux monochrome camera (right) during the growth of \ce{Li_2CuO_2} in 100 bar 80:20 \ce{Ar}:\ce{O_2}. (b) Laser power as a function of time during the crystal growth.}
\end{figure}

Crystal growth was performed within an 80:20 \ce{Ar}:\ce{O_2} atmosphere with a total pressure of 100~bar, which suppressed volatility.   A stable molten zone was maintained throughout using a growth rate of 10~mm/hr. In order to improve control of the laser power at the low power levels required for melting this compound (with a molten zone temperature near 1170\textdegree{}C), growth was performed by pulsing the laser output with frequency 100~Hz, a $50$\% duty cycle and total average power of 55~W. Images taken during the growth as well as the average growth power during growth are shown in Fig. \ref{fig:Li2CuO2_growth}.

\begin{figure}
\centering
\includegraphics[width=\columnwidth]{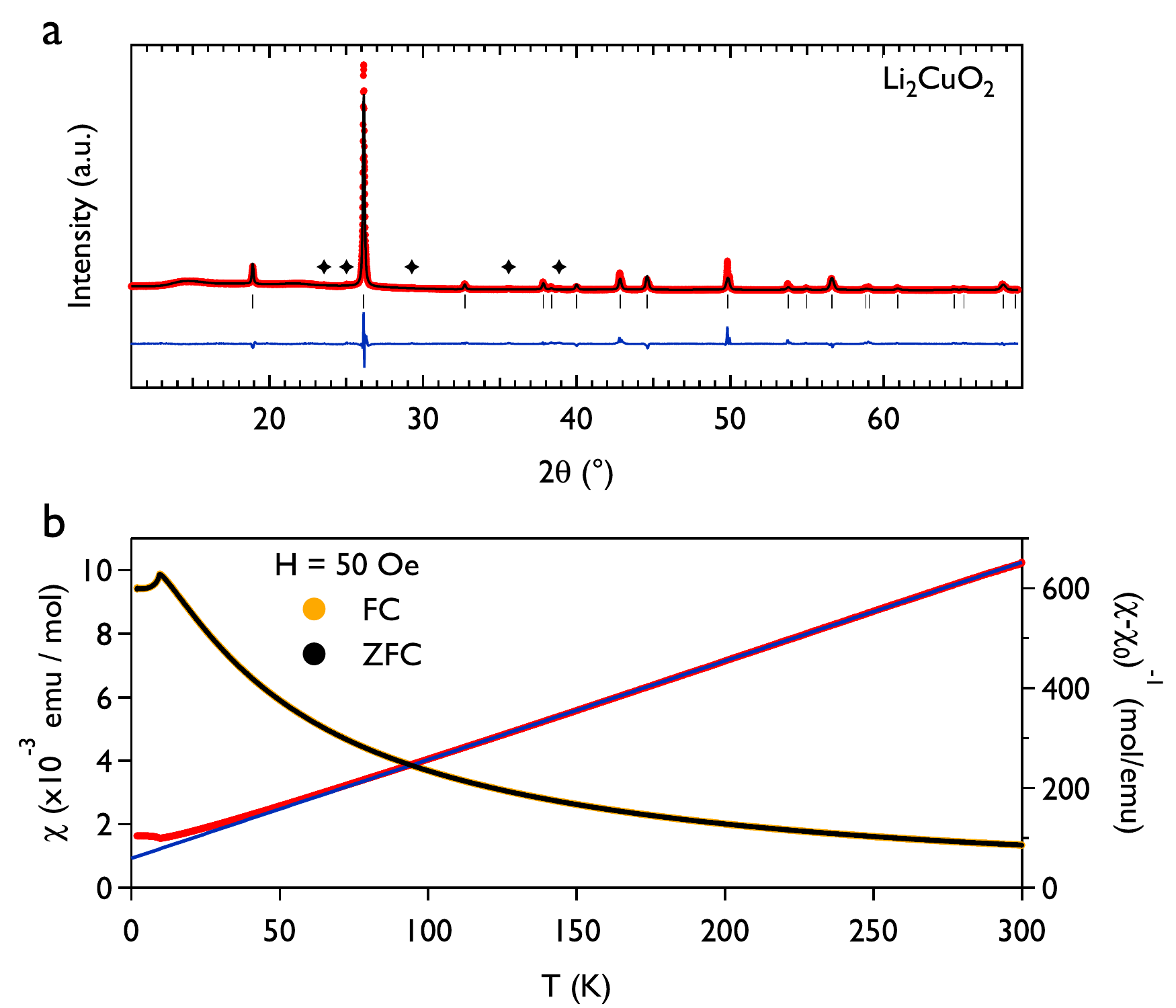}
\caption{\label{fig:Li2CuO2_figure}Characterization of the grown \ce{Li_2CuO_2} crystal. (a) Powder X-ray diffraction on a crushed piece of crystal. All observed peaks can be indexed to the reported \ce{Li_2CuO_2} structure (black tick marks) apart from small impurity peaks labelled with a star. (b) SQUID magnetometry on a 10.7~mg piece of the crystal. The Curie-Weiss fit to $1/\chi$ (blue line) yields $\theta_{CW}=-30$~K.}
\end{figure}

The resulting \ce{Li_2CuO_2} crystal is shown in Fig. \ref{fig:sample_images} (d) with a resolution-limited (101) facet characterized via x-ray diffraction. Inductively-coupled plasma, atomic emission analysis of the grown crystal revealed an atomic ratio of Li:Cu of 1.99:1.01.  The composition of the sample was confirmed by powder XRD on a piece of the crushed crystal (Fig. \ref{fig:Li2CuO2_figure}). XRD data refines to the expected structure for Li$_2$CuO$_2$ albeit with a considerable degree of preferred orientation ($R_p$ = 8.45\%, $\chi^2$ = 12.3) and a small percentage ($<$1\%) of an unknown impurity phase that appears to form predominantly at the crystal surface. Magnetization data are plotted in Fig. \ref{fig:Li2CuO2_figure} (b), and these data are consistent with previous reports of stoichiometric crystals \cite{shu_2017} showing no low-temperature ferromagnetism and suggest a near ideal oxygen stoichiometry.

\section{Summary}
We have developed a high-pressure, laser-heated floating zone furnace for the actively controlled crystal growth of high-purity, volatile compounds. The use of focused laser beams as the furnace heating source allows for extremely sharp heating gradients, while also allowing for the use of a solid metal growth chamber. This in turn greatly increases the pressures possible during growth, and a pressure of 1000 bar is achievable in our design. Successful growths of a wide range of oxide crystals was demonstrated and illustrate the performance capabilities of the instrument.  As a simple example of high pressure growth, crystals were grown under pressures up to 675~bar. While the examples presented here focused on the growth of a variety of complex oxides, we envision the HP-LFZ furnace will enable floating zone growth across a much broader array of volatile and metastable materials.

\begin{acknowledgments}
Funding support was provided by the W.M. Keck Foundation. We gratefully acknowledge fruitful discussions with David Bothman, Chris Torbet, and Tresa Pollock. The MRL Shared Experimental Facilities are supported by the MRSEC Program of the NSF under Award No. DMR 1720256; a member of the NSF-funded Materials Research Facilities Network. 
\end{acknowledgments}
\bibliography{bibtexfile}
\end{document}